\newcommand {\be}{\begin{equation}}
\newcommand {\ee}{\end{equation}}
\newcommand {\la}{\label}
\newcommand {\r}{\ref}
\newcommand {\bea}{\begin{eqnarray}}
\newcommand {\eea}{\end{eqnarray}}

\documentstyle[twocolumn,aps,prl]{revtex}
\draft
\begin{document}
\title{Origin of spin gap in CaV$_4$O$_9$: effect of frustration and
lattice distortion} 
\author{O. A. Starykh$^{1}$, M. E. Zhitomirsky$^{2}$, D. I. Khomskii$^{3}$,
 R. R. P. Singh$^{1}$, and K. Ueda$^{2}$}
\address{$^{1}$Department of Physics, University of California, Davis, 
California 95616\\
$^{2}$Institute for Solid State Physics, University of Tokyo,
Tokyo 106, Japan\\
$^{3}$Groningen University, 9747 AG  Groningen, The Netherlands}
\date{today}
\maketitle{}

\begin{abstract}
We study the origin of spin-gap in recently discovered
material CaV$_4$O$_9$.
We analyze the spin-$1/2$ Heisenberg model on the
$1/5$ depleted square lattice with nearest neighbor (nn)
and next nearest neighbor (nnn)
interactions,
in terms of the singlet and triplet states of the 4-spin
plaquettes and 2-spin dimers. 
Phase diagram of the model is obtained
within a linear ``spin-wave"-like approximation, and is shown to 
agree well with the earlier
results of QMC simulations for nn interactions.
We further propose that the special lattice structure of CaV$_4$O$_9$
naturally leads to lattice distortions, which enhances the
spin-gap via a spin-Peierls mechanism.
\end{abstract}

Recent discovery \cite{taniguchi} of a quantum disordered phase 
and spin-gap
in the layered
magnet CaV$_4$O$_9$  has attracted considerable interest
\cite{ueda,katoh,sano,mila,troyer}. The magnetic 
system can be described by a
Heisenberg model for spins of Vanadium ions ($S=1/2$)
on a 1/5-depleted square lattice.
At each site of this bipartite lattice three bonds meet: two of them 
belong to the 4-spin plaquettes covering the lattice (plaquette bonds),
whereas the third one (dimer bond) connects a plaquette with its 
neighbor (see Fig.~1). Since the coupling between spins is mediated by
superexchange via intermediate Oxygens, a strong 
next nearest neighbor interaction is also expected \cite{ueda,sano}.

We are thus led to the following Hamiltonian:
\begin{equation}
\hat{H} = \sum_{\text{n.n.}}J_{nn} {\bf S}_i\cdot{\bf S}_j  
+ J_2\sum_{\text{n.n.n.}}{\bf S}_i\cdot{\bf S}_j   \ ,
\label{H1}
\end{equation}
where the nn interaction $J_{nn}$ equals
$J_0~(J_1)$ for plaquette (dimer) bond.
It is evident that
this model has  disordered singlet 
ground states in two limits: 
for $J_0\gg J_1,J_2$
the ground state is a product of singlets on each plaquette, and
for $J_1\gg J_0,J_2$ it consists of singlets on the 
dimers.
However, a physically relevant choice of exchange parameters 
is $J_0\approx J_1$. 

For model~(\r{H1}) 
a spin-gap can arise either
from the non-equivalence of the dimer and plaquette bonds ($J_0
\neq J_1$),
or from the effects of frustration due to 
nnn couplings.
Previous theoretical works have mainly 
focused on the former.
Quantum Monte Carlo (QMC) simulations, 
done with $J_2=0$ \cite{troyer}, show that
the 1/5-depleted square lattice has Neel order
for $J_0=J_1$. However, the system
is close to the transition into the disordered phase.

The aim of this paper is twofold. 
First, we describe an efficient analytical
approach to study the $T=0$ phase diagram of (\r{H1}) in the
$J_1/J_0 - J_2/J_0$ plane, and show that a moderate 
$J_2 \sim 0.2J_0$, with $J_1=J_0$,
can account for the experimental data.
Second, we point out that the special lattice structure of this
system inevitably leads to lattice distortions, resulting in the
strengthening of the Plaquette bonds with respect to the Dimers.
Thus the spin-Peierls mechanism cooperates with
the intrinsic tendencies of the system in forming a spin-gap.

The phase diagram of the Hamiltonian (\ref{H1}) for
$J_2=0$ has been studied analytically using linear 
spin-wave \cite{ueda}, strong coupling perturbation \cite{ueda,katoh},
mean-field Schwinger boson \cite{mila} and cluster \cite{ueda} 
approaches. As expected, both spin-wave and Schwinger boson 
theories substantially overestimate the region of stability
of the N\'eel phase compared to QMC data \cite{troyer}, 
while expansions around strong-coupling limits
favor disordered phases. Here, we investigate this
model using bosonization techniques which correctly account for
 short-range spin correlations inside 
the plaquette and dimer blocks.

For the closely related case of the frustrated  
square-lattice antiferromagnet the disordered spin-state is
a dimer state, formation of which is
accompanied by a spontaneous breaking of the lattice symmetry
\cite{sachdev,gelfand}. In contrast, for the
CaV$_4$O$_9$-lattice the choice of spins which form singlets
is determined by the structure of the lattice and there is
no spontaneous symmetry breaking.
However, two different types of disordered short-range RVB states, 
with spin-singlets formed on plaquettes and dimers, are possible
\cite{ueda,katoh}.
For different values of model parameters one needs
to consider representations for spin operators in terms of
both dimer and plaquette states. We generalize previous derivations
of such representations \cite{sachdev,barab,blnchr,chub1} for the
two cases.

The starting point of these representations are 
noninteracting spin blocks. Let states of a single block
be given by $|\alpha\rangle$. In case of 
dimers, they are a singlet $|s\rangle$ 
($E_s=-\case{3}{4}J_1$) and a triplet 
$|t_\alpha\rangle$, $\alpha=x,y,$ and $z$ ($E_t=\case{1}{4}J_1$). 
All 16 states of a four spin plaquette 
can be found in Ref.~\cite{ueda,barab}. The lowest levels,
once again, are a singlet
with energy $E_s =-2J_0 + \case{1}{2}J_2$, and a triplet
with $E_t =-J_0 + \case{1}{2}J_2$. 
In the bosonic representation for plaquette spins 
we will restrict ourselves to
these four states only. This assumes 
that occupation numbers of all higher levels are small.

The site spins ${\bf S}_i$ are expressed in terms of the basis
block states as
\begin{equation}
{\bf S}_i = \langle\alpha| {\bf S}_i |\beta\rangle\,Z^{\alpha\beta}\ , 
\la{2}
\end{equation}
where $Z^{\alpha\beta}$ is the projection operator
$|\alpha\rangle\langle\beta|$ and
summation over repeated indices is assumed.

Consider first the matrix elements
that occur in Eq.(\r{2}), in the subspace of
one singlet and three triplet states. Using rotational
invariance in spin space and time-reversal symmetry
one gets
\bea
 & & \langle s|{\bf S}_i^\alpha|s\rangle = 0,  \ \ 
\langle s|{\bf S}_i^\alpha|t_\beta\rangle = 
\delta_{\alpha\beta} A_{st}^i, \
\nonumber\\ 
 & & \langle t_\alpha|{\bf S}_i^\beta|t_\gamma\rangle = 
i e^{\alpha\beta\gamma} A_{tt}^i , 
\eea
where $e^{\alpha\beta\gamma}$ is the totally antisymmetric tensor
and $A_{st}^i$, $A_{tt}^i$ are real.
Let each spin of the block be decomposed as
${\bf S}_i = {\bf L}_i + {\bf M}_i$,
where ${\bf L}_i$ has nonzero matrix elements between triplets
and singlet, while ${\bf M}_i$ acts between triplets only. It
is easy to see that ${\bf M}_i$ must be proportional
to the total spin of the block, and, thus, is independent of
$i$, whereas ${\bf L}_i\propto(-1)^i$ for blocks consisting
of equivalent spins.
Further calculations using explicit forms of singlet
and triplet states are straightforward and give
$A_{st}^i = (-1)^i/2,\ A_{tt}^i = 1/2$ for dimers,
and $A_{st}^i = (-1)^i/\sqrt{6},\ A_{tt}^i = 1/4$ for plaquettes.

We define the vacuum $|0\rangle$ and four boson 
operators that yield the four physical states
$|s\rangle=s^+|0\rangle$,  
$|t_\alpha\rangle=t_\alpha^+|0\rangle$.
The projection operators are naturally expressed as
$Z^{st_\alpha} = s^+t_\alpha$,   
$Z^{t_\alpha t_\beta} = t_\alpha^+t_\beta$, and so on.
Block spins represented via these boson operators are 
\bea
& & S_i^\alpha = \case{(-1)^i}{2}(s^+t_\alpha+t_\alpha^+s)
 - \case{i}{2} e^{\alpha\beta\gamma}t_\beta^+t_\gamma \ ,  
\ \ \text{for dimers}, 
\nonumber\\[-2mm]
\la{repres}
\\[-2mm]
& & S_i^\alpha = \case{(-1)^i}{\sqrt{6}}(s^+t_\alpha+t_\alpha^+s)
 - \case{i}{4} e^{\alpha\beta\gamma}t_\beta^+t_\gamma \ ,  
\ \ \text{for plaquettes}. \nonumber
\eea

Commutation relations between spins are satisfied
as long as bosonic representation preserves the algebra of 
the projection operators,
i.e. the relation
$Z^{\alpha \beta}\,Z^{\alpha' \beta'} =
\delta_{\alpha' \beta} Z^{\alpha\beta'}$.
This requirement restricts the number of bosons allowed on 
each block to one:
\begin{equation}
s^+s + t_\alpha^+t_\alpha = 1 \ .
\label{constraint}
\end{equation}
With the help of this constraint the Hamiltonian of a single block 
becomes
$\hat{H}_B = E_s s^+s + E_t t_\alpha^+t_\alpha$.
There are two ways to implement the constraint (\ref{constraint}).
First is the slave-boson treatment of Sachdev and Bhatt
\cite{sachdev}, 
where relation (\ref{constraint}) is imposed by a 
Lagrange multiplyer.
The second approach \cite{chub1} is to express the
$s$-operators in terms of the
$t_\alpha$-operators
\begin{equation}
s^+=s= \sqrt{1-t_\alpha^+t_\alpha} \ , 
\label{HP}
\end{equation}
and substitute into Eq.~(\ref{repres}).
Instead of the Holstein-Primakoff type
representation (\ref{HP}), we can also consider
an analog of the nonhermitian 
Dyson-Maleev representation: $s^+=1$ and $s=1-t_\alpha^+t_\alpha$ 
\cite{blnchr}.
Note that the block Hamiltonian is treated identically in both approaches.
Differences appear when interactions between the blocks are
switched on.

As in case of spin-wave theory
for ordered magnetic phases, one might expect even the
linear approximation, which neglects interaction between 
excitations, to work well.
This approximation can be formulated 
for both types of 
block spin representations in the same way. 
It consists of replacing $s$ and $s^+$ by 1, when calculating 
$({\bf S}_i\cdot{\bf S}_j)$ for pairs of spins from different blocks.
Also, only terms of second order in triplet operators should be kept.
Diagonalization of the resulting quadratic form 
is done by a standard Bogoliubov transformation, 
and one finds a $3$-fold degenerate spectrum of triplet excitations. 
For small coupling
between the blocks the spectrum is positive with a gap. 
Increasing the interaction between the blocks decreases the gap, 
which finally 
vanishes at the transition between disordered and ordered phases.

First, consider the plaquette singlet phase
which exists for large $J_0$. The spectrum of spin-1
excitations is threefold degenerate and has the 
dispersion
\begin{equation}
\omega_p^2({\bf k}) = J_0\left[J_0+\case{2}{3}(J_1-2J_2)
(\cos k_x+\cos k_y)\right] .
\label{plaq-fr}
\end{equation}
The minimum of the spectrum is at
$(\pi,\pi)$ for $(J_1-2J_2)>0$
and at $(0,0)$ for $(J_1-2J_2)<0$. From Eq.~(\ref{plaq-fr})
one finds the region of stability of the plaquette
phase, shown in Fig.~2. At $J_2=0$, 
singlets on 4-spin plaquettes become unstable at the critical ratio
$J_0/J_1|_{\text{cr}}=\case{4}{3}$, which is not too far from the
QMC estimate $J_0/J_1|_{\text{cr}}=1.1$ \cite{troyer}. 
The 
total energy of this phase, per spin, is 
\begin{equation}
E_{\text{g.s.}}^p  =  -\case{1}{2} J_0 + \case{1}{8} J_2 + 
\case{3}{8N}\sum_{\bf k} \left[\omega_p({\bf k})-J_0\right] \ . 
\label{plaq-en}
\end{equation}
It consists of the energy of noninteracting plaquette
singlets and the energy of zero-point fluctuations.

In the dimer state each crystal unit cell has two dimers.
Therefore, there are
two different branches of $S=1$ magnons 
in the Brillouin zone.
However, calculations  
are greatly simplified if instead we consider only one type
of dimers, which are defined in the new Brilloin zone corresponding
to the lattice formed by the centers of dimer bonds. This procedure
is quite similar to performing a spin-wave expansion around the
rotating quantization axis, as is often done in standard
spin-wave theory.
As a result, we obtain one triply degenerate excitation mode
in the new Brillouin zone,
which is twice the original one,
\begin{eqnarray}
\omega_d^2({\bf k}) & = & J_1\left[J_1-(J_0-J_2)
(\cos k_x-\cos k_y) \right. \nonumber \\
 & & \left. \mbox{}\ \ \  - J_2 \cos(k_x+k_y)\right] .
\label{dmr-fr}
\end{eqnarray}
The minimum of the spectrum is 
at ${\bf k}=(0,\pi)$ for $J_2 < \case{1}{3}J_0$,
and moves into the interior of the zone
for larger $J_2$.
At $J_2=0$, the dimer phase is unstable for 
$J_0/J_1>\case{1}{2}$, while the corresponding critical
ratio from QMC is 
$J_0/J_1|_{\text{cr}}=0.6\pm0.05$ \cite{troyer}. 
Note that the instabilities of the N\'eel
phase at $J_2=0$, as predicted by our linear bosonic theory, 
coincide with the mean field 
cluster approach \cite{ueda}. However, the latter is not appropriate
to study spin-liquid phases.
The total energy per spin in the dimer phase is 
\begin{equation}
E_{\text{g.s.}}^d  =  -\case{3}{8} J_1 + 
\case{3}{4N}\sum_{\bf k} \left[\omega_d({\bf k})-J_1\right]\ .
\label{dmr-en}
\end{equation}

Phase diagram of the model (\r{H1}) follows from 
Eqs.~(\r{plaq-fr}) and (\r{dmr-fr}). When 
frustration exceeds the critical value $J_2^{\text{cr}}=0.25J_0$,
the N\'eel phase ceases to exist. A first order transition line
separates plaquette and dimer phases 
since their symmetries are different \cite{ueda}.
This line is found by comparing 
(\r{plaq-en}) and (\r{dmr-en}) and is shown in Fig.~2
\cite{snoska}.

As $J_2$ grows, the energies of the omitted plaquette states 
decrease making the linear approximation less satisfactory, and at $J_2=J_0$
the second singlet which consists of two crossing dimers
becomes the ground state of the 4-spin plaquette \cite{ueda,barab}.
We have checked that this phase is not stable in the linear approximation
for any values of the parameters. Another possible short-range RVB state
is the plaquette-RVB on
large squares formed by $J_2$ bonds,
but it was also found to be unstable.
Therefore, for large values of $J_2$ magnetic order
should be stabilized again.
It is easy to see that in the limit $J_2 \gg J_0,J_1$ spins
are arranged into two interpenetrating N\'eel ordered sublattices which
are decoupled at the classical level. The degeneracy with respect to
a relative orientation of antiferromagnetic vectors
should be 
removed by quantum fluctuations, providing an example of
`order-from-disorder' phenomenon \cite{order}. We have presented in Fig.~2
transition line between this ordered state and disordered plaquette phase. 

 A theoretical comment is in order here. Calculations described
above can be repeated within an alternative slave-boson treatment
\cite{sachdev}, which amounts to introducing a singlet condensate
${\bar s}^2 \neq 1$ and a chemical potential $\mu$, and finding
their values self-consistently. Solving the self-consistent
equations for the phase boundaries, we found, at $J_2=0$, that dimer 
phase is stable
for $J_0/J_1 \leq 0.874$, whereas plaquette-RVB exists up to $J_0/J_1 =0.76$.
 Thus, unlike linear approximation, mean-field
slave-boson treatment predicts $no$ ordered phase even at $J_2=0$, in
striking contradiction with numerical simulations and 
linear approximation results. Situation is not improved even upon taking
triplet-triplet interaction into account - its effect on the location
of phase boundaries is ridiculously small, less then $1$ per cent.
These findings clearly favor our linear approximation as the most
suitable one for the problem at hand.

To make contact with experiments, let us 
first confine ourselves to $J_0=J_1$.
Note that only second terms
of Eq.~(\r{repres}) contribute to the uniform susceptibility,
leading to 
$\chi \sim \case{1}{TN}\sum_{\bf k} n(\omega_{\bf k})(1 + n(\omega_{\bf k}))$,
where $n(\omega)$ is the Bose factor. We can compare the Curie-Weiss
relation $\theta=\case{3}{4}(J_0+J_2)$ and the gap $\Delta$, in 
Eq.~(\r{plaq-fr}) appropriate for the PRVB phase, with the
experimental Curie Weiss constant 
and the gap determined from the T-dependence of the susceptibility.
We find that the experimental numbers $\Delta=107$~K and
$\theta=220$~K \cite{taniguchi}
can arise from two sets of exchange constants
(i) $J_0=245~K$, $J_2=48~K$ and (ii) $J_0=193~K$, $J_2=100~K$.
Thus this theory, by itself, can provide a consistent
description of the properties of CaV$_4$O$_9$ observed so far.

However, we would now like to point out an additional aspect of
this material which has so far gone unnoticed. Because of the
special lattice structure, spin-phonon couplings will cause a
gain in magnetic energy which is {\bf linear} in the lattice
distortion, whereas the loss in elastic energy is always quadratic.
Thus there will always be a lattice distortion in this system
which will cause the plaquette bonds to shrink and the dimer
bonds to elongate. This will lead to $J_0>J_1$. This spin-Peierls
mechanism will enhance the stability of the PRVB phase and
contribute to the spin-gap.

To get a quantitative  estimate, let the shrinking of the 
plaquettes change the distance between the Vanadium ions from
$R$ to $R\pm 2u$. Using the phenomenological relation $J(R)\approx
const/R^{10}$ \cite{harrison}, the exchange constants $J_0$ ($J_1$)
increase (decrease) by
$\delta J = 20uJ/R$. 
The magnetic energy gain per Vanadium atom is $e_m= -\delta J(C_1 - 0.5 C_2)$,
where $C_1$ and $C_2$ are spin correlations on the plaquette and
dimer bonds in the absence of distortion. 
It is sufficient to evaluate them on the Neel state, where
they both are $-1/4$.
Thus the gain in magnetic
energy per unit volume is
$E_m \sim (20uJ/8R) (1/R^2 R_{\perp})$,
where $R_{\perp}$ is the distance between magnetic layers.
A rough estimate for the elastic energy per unit volume is
$E_{\text{ph}} \sim \case{1}{2} B (u/R)^2 $,
where $B$ is a bulk modulus of the material. Minimizing
the energies leads to a distortion
$u \sim (5J/2BRR_{\perp})$, which induces a variation
of the exchange integrals by
\be
\delta J/J \approx (50J/R^2 R_{\perp} B).
\la{variation}
\ee
Taking $R=3$~\AA, $R_{\perp}=5$~\AA, $J=300$~K, 
$B=10^{12}$~dynes/cm$^2$, we get
$u/R \sim 10^{-3}$ and $\delta J/J\sim 10^{-2}$--$10^{-1}$.
These values are similar to what is found in
most of the spin-Peierls materials \cite{review}, where
$J_0/J_1 \sim 1.3$. This similarity is not unexpected
as the linear gain in magnetic energy
is somewhat analogous to $u^{4/3}$ gain in a 1D spin-$\case{1}{2}$ 
chain \cite{cross}.
 
If $J_2$ was negligible, and the spin-Peirels mechanism was
entirely responsible for the spin-gap, we would use
Eq.~(\r{plaq-fr}) with
$J_2=0$ and the Curie-Weiss relation
$\theta=\case{1}{4} (2J_0 + J_1)$ to compare with experiments. 
This leads to estimates
$J_0=331$~K, $J_1=218$~K. The ratio 
$J_0/J_1 \sim 1.5$ corresponds to
$\delta J =0.2 J$, which is somewhat larger than our earlier back of
the envelop estimate, suggesting that some $J_2$ is
needed to explain the data.

We note that here the spin-Peierls mechanism
does not lead to a spontaneous breaking of the lattice symmetry.
Hence, we do not expect a sharp
change in the lattice distortion $u$ with temperature.
Instead, it should follow short-range order and
the spin-gap should be temperature-dependent.
Finally, we point out that
most known spin-Peierls materials have first order 
structural phase transition at high temperature \cite{review}. 
This creates a soft
phonon mode in the system, which in the end favors the spin-Peierls
phenomena \cite{bulaev}. Curiously enough, a small discontinuous jump
in $\chi(T)$  at $340$~K was observed also in CaV$_4$O$_9$ 
\cite{taniguchi},
but was attributed to an admixture of the VO$_2$
phase. In view of our proposal this point should be studied more
carefully.

 Our discussion of possible spin-Peierls phenomena is only an 
order of magnitude estimate. Nevertheless it shows that
spin-phonon coupling may successfully ``cooperate'' with intrinsic tendencies
for forming a spin-gap due to the frustrating next-nearest-neighbor
interaction present in CaV$_4$O$_9$.

OAS acknowlegdes helpful conversations with A. V. Chubukov.
Two of us (DIK and RRPS) would like to thank the
Gordon Godfrey Foundation for support at the University of
New South Wales, where part of this work was done. OAS and RRPS
are supported in part by NSF grant number DMR-9318537.
DIK was supported in part by FOM, the Netherlands.

\begin{figure}
\caption
{Lattice structure of CaV$_4$O$_9$. Three types of exchange
bonds are indicated by thick lines. The pattern of lattice
distortion is shown schematically by thin dashed lines.}
\label{fig.1}
\end{figure}

\begin{figure}
\caption
{Phase diagram of the model in linear approximation.
Thick (thin) solid line denotes second (first) order phase transition.
Regions of stability of dimer (plaquette) phase are shown by long
(short) dashed lines.}
\label{fig.2}
\end{figure}


\begin{references}
\bibitem{snoska} small deviation of the first order transition line
from the crossing point of two instability lines is an artifact 
of the linear approximation and can be considered as a `systematic error
bar' put on this line.
\end{references}
\end{document}